\documentclass{ws-procs9x6}

\usepackage[dvips]{color} 
\usepackage{epsfig}
   
\newcommand{\beq}{\begin{equation}}
\newcommand{\eeq}{\end{equation}}
\newcommand{\bq}{\begin{equation}}
\newcommand{\eq}{\end{equation}}
\newcommand{\ba}{\begin{array}}
\newcommand{\ea}{\end{array}}
\newcommand{\beqa}{\begin{eqnarray}}
\newcommand{\eeqa}{\end{eqnarray}}

\newcommand{\ZZ}{\mathbb{Z}}
\newcommand{\SB}{\mathbb{S}}

\def\TT{{\mathcal T}}

\def\End{\end{document}}

\def\to{\rightarrow}

\def\dis{\displaystyle}
\def\f{\frac}
\def\ov{\overline}
\def\[{\left[}
\def\]{\right]}
\def\({\left(}
\def\){\right)}

\def\U1EM{U(1)_{\rm em}}

\def\leqq{\leqslant}
\def\geqq{\geqslant}

\def\KF{{\color{red}\kappa^4}}
\def\KT{{\color{red}\kappa^2}}
\def\KBT{{\color{red}{\mathbf\kappa}^2}}

\def\[{\left[}
\def\]{\right]}
\def\dis{\displaystyle}

\def\cut{\Lambda}
\def\Ucut{\Lambda_{\rm U}}

\def\d{\delta}

\def\gh{\widehat{g}}

\def\Ah{\widehat{A}}

\def\NKK{N_{\rm KK}}

\def\ep{\epsilon}

\begin{document}

\title{Unitarity in Higher Dimensions \\ 
         and Gauge Unification
\footnote{\uppercase{T}alk presented by 
\uppercase{H}ong\uppercase{-J}ian \uppercase{H}e
at {\it \uppercase{SUSY} 2003:
\uppercase{S}upersymmetry in the \uppercase{D}esert}\/, 
held at the \uppercase{U}niversity of \uppercase{A}rizona,
\uppercase{T}ucson, \uppercase{AZ}, \uppercase{USA}, 
\uppercase{J}une 5-10, 2003.     }}        


\author{R. Sekhar Chivukula\,$^1$,~~ Duane A. Dicus\,$^2$, \\ 
        Hong-Jian He\,$^2$, 
        ~~\lowercase{and}~~ 
        S. Nandi\,$^3$}

\address{
\vspace*{1mm}
$^{1}$\,Department of Physics, Boston University,
        Boston, MA 02215, USA, and\\[1mm]
        Department of Physics and Astronomy,
        Michigan State University, \\                                      
        3243 Biomedical Physical Sciences,
        East Lansing, MI 48824, USA\\[2mm]
$^{2}$\,Center for Particle Physics, 
      University of Texas at Austin, TX 78712, USA\\[2mm]
$^{3}$\,Department of Physics, Oklahoma State University,
      Stillwater,\,OK\,74078,~USA   
}


\maketitle

\abstracts{%
Unitarity of the 4d standard model is 
ensured by the conventional Higgs mechanism
with a fundamental spin-0 Higgs boson, responsible for  
gauge boson mass-generations.
On the contrary, Kaluza-Klein (KK) compactification of 
extra spatial dimensions
can geometrically realize the gauge boson mass generation 
without invoking a fundamental Higgs scalar. We reveal that 
massive gauge boson scattering in the compactified theories is unitary at 
low energies, and the unitarity violation is {\it delayed} 
to the intrinsic ultraviolet (UV) 
scale of the higher dimensional gauge theory. 
We demonstrate that 
this is a generic consequence of the 
``geometric Higgs mechanism'' (GHM), 
manifested via Kaluza-Klein equivalence theorem (KK-ET). 
We further show that 
the presence of many gauge KK states below the UV cutoff scale  
imposes strong bounds on the highest KK level ($\NKK$). 
Applying these bounds to higher-dimensional SUSY GUTs implies 
that only a small number of KK states can be used to accelerate gauge
coupling unification, and suggests that the GUT scale 
in the 5d minimal SUSY $SU(5)$ is above $10^{14}$\,GeV.%
}

\section{The Puzzle: Unitarity in 4d versus Higher-d}

Artists have explored extra dimensions since at least 1909\,\cite{picasso},
more than a decade before the first scientific 5d theory proposed by 
physicists Kaluza-Klein (KK)\,\cite{KK}
who attempted to unify electromagnetism with
Einstein gravity.
When gauge bosons propagate in higher dimensional space, 
the compactification
results in KK towers of {\it massive} vector bosons in 4d, with
their masses characterized by the (inverse) size $R$ of the extra dimensions.
This provides a ``geometric'' realization 
of gauge boson mass generation, contrary
to the conventional Higgs mechanism which invokes a fundamental
spin-0 physical Higgs boson\,\cite{Higgs}.

A Higgsless Yang-Mills gauge theory 
can have a gauge boson mass term put in by hand,
while perfectly respecting the gauge symmetry in the nonlinear realization. 
The real problem is {\it Unitarity Violation}, i.e., the scattering of the
longitudinal components of the massive gauge bosons 
has bad high energy behavior
because the longitudinal polarization vector grows with energy, 
$\,\ep_L^\mu (k) = \dis\f{k^\mu}{m_v} + O\!\(\f{m_v}{E}\)$,\, 
where $m_v$ is the gauge boson mass.  
In consequence, a Higgsless standard model (SM) 
can only be an {\it effective theory}
valid up to energies no higher than the unitarity violation 
scale,\cite{SM-UB}
$
E \,<\, \Ucut = \sqrt{8\pi}\,v \simeq 1.2\,{\rm TeV}.\,
$
As we know, in the 4d SM it is the inclusion 
of a {\it physical} Higgs boson that 
cures the bad high energy behavior.

Now it is natural to ask: 
{\it Why would we expect the higher dimensional gauge theory to be unitary?}
Consider first a non-compactified massless Yang-Mills theory 
in $D=4+\d$ dimensions, which is nonrenormalizable
because the gauge coupling has mass-dimension 
\,${\rm dim}(\widehat{g})=-\dis\d/2$.\, 
Although the $2\to 2$ scattering amplitude of $D$-dimensional 
{\it massless} gauge bosons behaves as constant of $O(\widehat{g}^2)$,
we note\,\cite{5d1,5d3}
that the $s$-partial wave grows with energy $\sqrt{s}=E$ due to the 
$D$-dimensional phase space
and is given by, in the $SU(k)$
gauge-singlet channel,\,\cite{5d1,5d3} 
\beq
\label{eq:a0-D}
\ba{l}
\hspace*{-2mm}
\widehat{a}_{00}^{~}  = \dis
\f{E^{\d}}{2(16\pi)^{1+\f{\d}{2}}\,\Gamma\!\(1\!+\!\f{\d}{2}\)}\!\!
\int^\pi_0 \!\!\! d\theta\,(\sin\theta)^{1+\d} \,\widehat{T}_0
= \dis 
\f{2k\sqrt{\pi}\[4(2+\d)-\f{\d^3}{1+\d}\]\!
(\widehat{g}\,E^{\f{\d}{2}}_{~})}
  {(16\pi)^{1+\d/2}\d(2\!+\!\d)\Gamma\!\(\f{1+\d}{2}\)}
. 
\ea
\eeq
Consequently, we find\,\cite{5d1,5d3} 
that the unitarity is violated at the intrinsic ultraviolet
(UV) scale of \,$O(\widehat{g}^{\,-2/\d})$\,,
\beq
\label{eq:UB-DSUk}
E ~\,<~\, \Ucut = \dis
\[\f{\dis(16\pi)^{1+\f{\d}{2}}\d(2+\d)\,\Gamma\!\(\f{1+\d}{2}\)}
    {\dis2k\sqrt{\pi}
     \left|4(2+\d)-\f{\d^3}{1+\d}\right|} \f{1}{\,\widehat{g}^2\,}
\]^{1/\d}.
\eeq

A realistic higher-dimensional theory will be compactified. 
{\it Why would a compactified gauge theory be unitary?} 
Note that the compactified gauge theory in 4d will
contain towers of spin-1 {\it massive KK gauge bosons}, 
whose masses are generated by the ``geometry'' 
rather than by a fundamental Higgs boson.
As explained earlier, the $W_LW_L$ scattering 
in the 4d SM is {\it non-unitary} 
without the Higgs contribution!
So, the {\it puzzle} is: 
{\it why would such a compactified KK gauge theory be
unitary at all\,?}

\section{\hspace*{-1.5mm}Geometric\,Higgs\,Mechanism\,\&\,Unitarity\,of\,5d\,Yang-Mills}

We start by considering a generic 5d Yang-Mills theory of $SU(k)$, compactified
on the orbifold $\SB^1/\ZZ_2$. This corresponds to imposing the Neumann and
Dirichlet boundary conditions (BCs)
on a line segment $[0,\,L]$ ($L\equiv\pi R$),
for the 5d gauge fields $(\Ah^a_\mu,\,\Ah^a_5)$ respectively,
\beq
\label{eq:BC1}
\left. \partial_5\Ah^a_\mu \right|_{x^5=0,L} =0\,,~~~~~
\left.          \Ah^a_5   \right|_{x^5=0,L} =0\,.
\eeq
%
Eq.\,(\ref{eq:BC1}) 
preserves the gauge symmetry $SU(k)$ in 4d, 
but we can construct other consistent BCs 
which reduces the rank of the gauge group, and in the simplest
case we fully break $SU(k)$ by the 
following BCs,\cite{5deMoose,5dgEWSB,xTerning}
\beq
\label{eq:BC2}
\left.\partial_5\Ah^a_\mu \right|_{x^5=0} \!=0\,,~~
\left. \Ah^a_\mu \right|_{x^5=L} \!=0\,;~~~\,
\left. \Ah^a_5   \right|_{x^5=0} \!=0\,,~~
\left.\partial_5\Ah^a_5   \right|_{x^5=L} \!=0\,.~~
\eeq
Thus we can derive Fourier expansions for  $(\Ah^a_\mu,\,\Ah^a_5)$,
and integrate out $x^5$. 
Under (\ref{eq:BC1}), the resulting 4d KK Lagrangian contains the
kinetic term,
\beq
\label{eq:KE}
\hspace*{-2mm}
{\mathcal L}_{\rm KE}^{~} = -{\f{1}{4}} \!
\left[\(\partial_{[\mu} A^{a0}_{\nu ]}\)^2
\!+\! \sum_{n=1}^{\infty} \(\partial_{[\mu} A^{an}_{\nu ]}\)^2 \right]
-{\f{1}{2}} \sum_{n=1}^\infty \!\left[M_n A^{an}_\mu -
\partial_\mu A^{an}_5\right]^2,
\eeq
where \,$M_n=\dis\f{n}{R}$\, is the mass of the KK state at level-$n$.
It is important to note that (\ref{eq:KE}) contains a mixing term
$\,M_n A_\mu^{an}\partial^\mu A_5^{an}$\,,\, which enforces the
dynamical conversion \,$A^{an}_5\Longleftrightarrow A_L^{an}$\, so that
each KK state $A_\mu^{an}$ acquires a longitudinal component and
becomes massive. Without invoking any extra {\it physical} Higgs boson,
this is a {\it geometric} realization of gauge boson
mass generation, which we call the ``Geometric Higgs Mechanism'' (GHM)
and whose important consequences will be explored below (\ref{eq:KK-ET1}).
To eliminate the $A^{an}_\mu-A_5^{an}$ mixing 
in (\ref{eq:KE}), we construct 
the general $R_\xi$ gauge-fixing term,\cite{5d1}
\beq
{\mathcal L}_{\rm GF}^{~} \,=\, \dis\sum_{n=0}^\infty 
-\f{1}{2\xi} \( F^{an}\)^2\,,~~~~
F^{an}= \partial^\mu A^{an}_\mu 
+ \xi M_n A^{an}_5 \,,
\eeq
where $\xi$ is the gauge-fixing parameter.
The Faddeev-Popov ghost term 
${\mathcal L}_{\rm FP}^{~}$ can be derived accordingly.
Without physical Higgs boson one may naively expect,
from the lesson of 4d Higgsless SM, that
the scattering of longitudinal KK gauge bosons would violate
unitarity at a scale
%
$~\dis\Ucut ~\sim~ \f{4\pi M_n}{g} 
= \f{4n \pi}{g R} 
= \f{4n\pi^{\f{3}{2}}}{\gh\sqrt{R}}                              
= \f{4n\pi^2 g}{\gh^2}\,,\,
$
%
where $g=\gh/\sqrt{\pi R}$. 
However, this cannot be true because by adjusting $R$
to be arbitrarily large, 
the unitarity violation scale $\Ucut$ would be
arbitrarily below the intrinsic UV scale of \,$O(1/\gh^2)$.\,
A deeper reason is yet to be sought!
\begin{figure}[h]
\label{fig:KK4n}
\vspace*{-8mm}
\hspace*{-8mm}
\includegraphics[width=13cm,height=4.8cm]{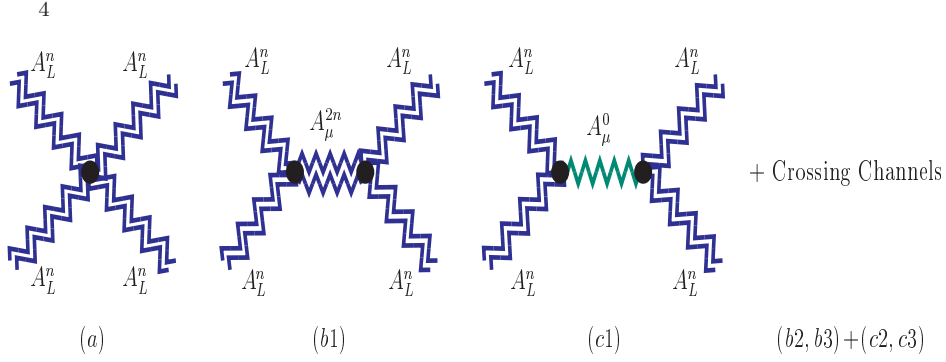} 
\vspace*{-8mm}
\caption{Longitudinal scattering
         $A_L^{an}A_L^{an}\to A_L^{an}A_L^{an}$
         in compactified KK theory.}
\end{figure}
%
We first compute the longitudinal scattering
$A_L^{an}A_L^{an}\to A_L^{an}A_L^{an}$ in Fig.\,1 
and find the leading amplitude is only of $O(E^0)$,\,\cite{5d1}
\beq
\label{eq:KKnnnn}
\ba{l}
\TT\!\[A^{an}_L A^{bn}_L \!\to\! A^{cn}_L A^{dn}_L\]  
~=~
g^2\!\[
C^{abe}C^{cde} T_1 +
C^{ace}C^{dbe} T_2 + 
C^{ade}C^{bce} T_3
\],
\\[2mm]
T_1= \dis \f{5}{2}c \,,~~~~
T_2= \dis -\f{~8c^2-5c+9~}{2(1-c)} \,,~~~~ 
T_3= \dis  \f{~8c^2+5c+9~}{2(1+c)}  \,,
\\[-3mm]
\ea
\eeq
where \,$c=\cos\theta$,\, 
and the small terms of \,$O(M_n^2/E^2)$\, are ignored.
%
$$
\ba{cccc}
\hline\hline
& & &
\\[-2mm]
\color{black}     {\rm ~~Diagram~~}      &
\dis\color{black}   T_1                    &
\dis\color{black}   T_2                    &
\dis\color{black}   T_3
\\[2mm]
\hline
& & &
\\[-2mm]
{\rm (a)}                                  &
\color{blue} ~~6c(\KF -\KT )~~                &
\color{blue} ~~\f{3}{2}(3-2c-c^2)\KF~~         &
\color{blue} ~~-\f{3}{2}(3+2c-c^2)\KF~~
\\[1mm]
                                           &
                                           &
\color{blue}  -3(1-c)\KT                   &
\color{blue}  +3(1+c)\KT
\\[1mm]
{\rm (b1\!\oplus\!2\!\oplus\!3)}           &
\color{blue}  -2c(\KF +\KT )               &
\color{blue}  -\f{1}{2}(3-2c-c^2)\KF       &
\color{blue}   \f{1}{2}(3+2c-c^2)\KF                                                  
\\[1mm]
                                           &
                                           &
\color{blue}  +3(1-c)\KT                   &
\color{blue}  -3(1+c)\KT                                                  
\\[1mm]       
{\rm (c1\!\oplus\!2\!\oplus\!3)}           &
\color{blue}  -4c\,\KF                     &
\color{blue}  (-3+2c+c^2)\KF               &
\color{blue}  (3+2c-c^2)\KF                                                  
\\[1mm]
                                           &
                                           &
\color{blue}  -8c\,\KT                       &
\color{blue}  -8c\,\KT         
\\[1mm]
\color{red} ~~{\rm\bf Sum}                  &
\color{red} {\bf -8c\KBT}                   &
\color{red} {\bf -8c\KBT}                   &
\color{red} {\bf -8c\KBT}    
\\[1mm]
\hline\hline
\\[-3.3mm]
\ea
$$
~~~\,In the above Table, 
we summarize the nontrivial $O(E^4)$ and $O(E^2)$ 
cancellations,  with \,$\kappa\equiv E/(2M_n)$\, and \,$E\equiv\sqrt{s}$,
where the summed $O(E^2)$ terms are found to be exactly vanishing
after using Jacobi identity
$\,C^{abe}C^{cde} + C^{ace}C^{dbe} + C^{ade}C^{bce} = 0$\,.\,
We also verified exact $E$-cancellations in all  
other channels\,\cite{5d1,5d3}.
Hence, so long as the 4d gauge coupling is perturbative,
the individual KK scattering channel is manifestly unitary!

The Geometric Higgs Mechanism (GHM) we observe essentially results
from the 5d gauge symmetry and the proper compactification\,\cite{5d1}.
Based on the 5d gauge symmetry (or the equivalent BRST invariance),
we have extended the 4d derivation\cite{He-4d} to deduce
a 5d Slavnov-Taylor identity\cite{5d1,5deMoose},
\beq
\label{eq:KK-ET0}
\left< 0\right|T
\widehat{F}^{a_1^{~}}(\hat{x}_1^{~})
\widehat{F}^{a_2^{~}}(\hat{x}_2^{~})
\cdots
\widehat{F}^{a_N^{~}}(\hat{x}_N^{~})\,
\widehat{\Phi}\left|0\right> \,=\,0 \,,
\eeq
where 
\,$\widehat{F}^{a}= \partial^\mu\widehat{A}^a_\mu +
                 \xi\partial_5\widehat{A}_5^a $\, 
is the 5d gauge-fixing function, and $\widehat{\Phi}$ denotes
other possible amputated (non-gauge) physical fields in 5d.
After KK-expansion and amputation of the external fields
in $\widehat{F}^a$'s, we arrive at
\beq
\label{eq:KK-ET1}
\left< 0\right|
\ov{F}^{a_1^{~}n_1^{~}}(k_1^{~})
\ov{F}^{a_2^{~}n_2^{~}}(k_2^{~})
\cdots
\ov{F}^{a_N^{~}n_N^{~}}(k_N^{~})\,
{\Phi}\left|0\right> \,=\,0 \,,
\eeq
where 
\,$\ov{F}^{an}(k)
= ik^\mu {A}^{an}_\mu - C^{an}M_n^a{A}_5^{an} 
= iM_n^a \(A_S^{an}+iC^{an}A_5^{an}\)
$\, with 
$A_S^{an}=\ep^\mu_S A^{an}_\mu$  
($\ep^\mu_S \equiv k^\mu/M_n^a$)
and
$\,C^{an}=1 +O({\rm loop})$\,\cite{He-4d}.\, 
Our identity (\ref{eq:KK-ET1}) just shows that the 
unphysical scalar-KK-component
($A_S^{an}$) and the 5th gauge-KK-component $A_5^{an}$ are
{\it confined} at the $S$-matrix level, so they together have
zero contribution to any physical process. This is nothing
but a quantitative formulation of the 
{\it Geometric Higgs Mechanism} (GHM) at the
$S$-matrix level, where 
{\it $A_5^{an}$ serves as
the geometric would-be Goldstone boson and 
gets converted to the physical longitudinal component $A_L^{an}$
at each KK-level.}
The implications of this GHM are profound: 
{\bf (i)} it ensures\cite{5d1} the nontrivial $E$-cancellations 
and the unitarity of $A_L^{an}A_L^{am}$ scatterings;
{\bf (ii)} it suggests\cite{5d1,5d2,5d3} the possibility that
the electroweak symmetry breaking may be realized 
{\it geometrically} without physical Higgs boson\cite{5deMoose,5dgEWSB}.
Noting that the $\Ah^a_L-\Ah^a_5$ conversion occurs under the
5d compactification, 
we require that the consistent BCs be imposed such that
the action respects the 5d gauge symmetry 
(or the equivalent 5d BRST invariance) at the boundaries.
We observe\,\cite{5deMoose,5dgEWSB} that 
all such consistent BCs can be reconstructed and classified from 
the continuum limits of proper gauge-invariant lattice formulation
(deconstruction) of the extra dimensions.

Expanding the polarization vector 
$\ep^\mu_L=\f{k^\mu}{M_n^a}+O\!\(\f{M_n^a}{E}\)$, 
we have deduced, from the identity (\ref{eq:KK-ET1}),
the KK Equivalence Theorem (KK-ET)\cite{5d1},
\beq
\label{eq:KK-ET}
\dis
\hspace*{-3mm}
\TT \[A_L^{a_1n_1}, 
       A_L^{a_2n_2}, 
       \cdots \!, \Phi
\] ~=~ C_{\rm mod} 
\TT \[A_5^{a_1n_1}, 
       A_5^{a_2n_2}, 
       \cdots \!, \Phi
\] + O\!\(\! {M_{n}^{a}}/{E}\) ,
\eeq
where \,$C_{\rm mod} = (-i)^N [1+ O({\rm loop})]$.\,
We stress that our above formulation of KK-ET (\ref{eq:KK-ET})\cite{5d1} 
is completely general, {\it valid independent of whether
the gauge group rank is preserved or reduced under 
compactification, i.e., independent of whether the zero-mode
gauge fields remain massless or acquire KK-masses (or masses
induced by usual Higgs).}
A crucial {\it general} observation\cite{5d1} 
is that the KK-ET ensures the 
nontrivial $E$-cancellations in the $A_L^{an}$-amplitude,
as enforced by the GHM, 
because the $A_5^{an}$-amplitude on the RHS of 
(\ref{eq:KK-ET}) has all individual diagrams manifestly 
of $O(E^0)$ or smaller. 
Hence, in the $A_L^{an}$-amplitude all terms
of $O(E^q)$ ($q> 0$) must cancel down to $O(E^0)$.
\begin{figure}[h]
\label{fig:KK5-4n}
\vspace*{-11mm}
\hspace*{-15mm}
\includegraphics[width=14cm,height=5cm]{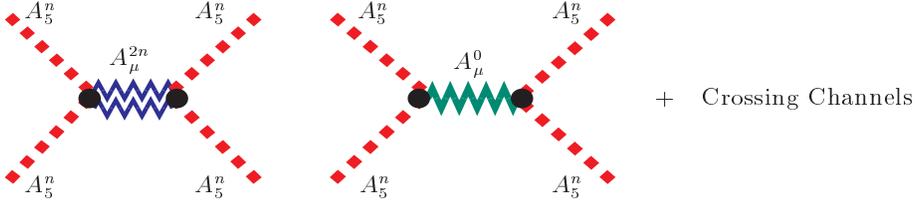} 
\vspace*{-17mm}
\caption{Goldstone scattering $A_5^{an}A_5^{an}\to A_5^{an}A_5^{an}$
in compactified KK theory.}
\vspace*{-2mm}
\end{figure}
Our direct computation of the $A_5^{an}$-scattering in Fig.\,\ref{fig:KK5-4n}
gives,\cite{5d1}
\beq
\label{eq:KK5nnnn}
\ba{l}
T\!\[A^{an}_5 A^{bn}_5 \!\to\! A^{cn}_5 A^{dn}_5\]  
~=~ 
g^2\!\[
C^{abe}C^{cde} \widetilde{T}_1 +
C^{ace}C^{dbe} \widetilde{T}_2 +
C^{ade}C^{bce} \widetilde{T}_3
\],
\\[2mm]
\widetilde{T}_1= \dis -\f{3}{2}c \,,~~~~
\widetilde{T}_2= \dis -\f{~3(3+c)~}{2(1-c)} \,,~~~~ 
\widetilde{T}_3= \dis  \f{~3(3-c)~}{2(1+c)}  \,,
%
\ea
\eeq
which is indeed {\it equivalent} to Eq.\,(\ref{eq:KKnnnn}) after applying
the Jacobi identity.

So far we have fully understood why the compactified higher-d
gauge theory does exhibit low energy unitarity in all individual
scattering channels. But, compactified KK theory is nonrenormalizable,
and we observe\cite{5d1} 
that the KK scattering has to reflect the bad 5d high energy
behavior in (\ref{eq:a0-D}) via coupled channels 
because of a large number of KK states existing below the UV cutoff 
\,$\cut =\NKK/R$\,.\,  In the gauge-singlet channel, 
$\left|\Psi\right>=\f{1}{\sqrt{N_0}}\sum_{\ell=0}^{N_0}
 \left|A_L^{a\ell}A_L^{a\ell}\right>$, 
we find that the unitarity condition 
indeed cuts off the KK tower at $N_0=N$ such that 
\beq
\dis\f{N}{R} \,~\lesssim~\, 
\f{\sqrt{32\pi}}{k}\f{\,O(1)\,}{\gh^2} 
 = O\!\(\f{1}{\,\gh^2}\),~
\eeq
reproducing the feature in (\ref{eq:UB-DSUk}).
The unitarity in the deconstructed 5d gauge theory was first studied
in Ref.\,[\cite{5d2}]. 
Our approach to the geometric electroweak symmetry breaking (GEWSB) 
is discussed elsewhere\cite{5deMoose,5dgEWSB}.

\section{Unitarity and Higher-d Gauge Unification}  

Extending the above generic results to the 5d SM, 
we derive\cite{5d3} strong limits
on the highest KK-level $\NKK$, as well as the zero-mode Higgs mass 
$m_H^{~}$\,,

\beq
\ba{lll}
&& \\[-9mm]
\hspace*{-4mm}
{\rm 5d\,QCD\!:}~~  
&  \NKK ~\leqq~ 4 \,,~~~~~ &  ({\rm for}~ \alpha_s\simeq 0.1)\,;
\\[0mm]
\hspace*{-4mm}
{\rm 5d\,EW\!:}~~ 
&  \NKK ~\leqq~ 11 \,,~~~~~ &  ({\rm for}~ g=2m_w^{~}/v)\,;
\\[0mm] 
&\dis  
~m_H^{~}\, ~<~ v\sqrt{16\pi/3}\,N_{\rm KK}^{-1/2} \simeq 303\,{\rm GeV},~\,\, 
&
({\rm for}~ \NKK = 11)\,.
\ea
\eeq

\vspace*{-11mm}
\begin{math}
\hspace{-\parindent}\hbox{\parbox{35mm}{\rm 
~~~With all gauge bosons propagating in the bulk, 
many KK states will contribute to \,the gauge\, coupling \\
running~and~accelerate\\ 
the~gauge~unification\cite{powerlaw}.\\ 
We apply the unitarity analysis to the 
minimal 5d SUSY GUT\cite{powerlaw} and 
find that imposing the unitarity limit 
$N_{\rm KK}^{EW}\leqq 11$~\,suggests\,~the~GUT\\
scale\,~${M_{\rm G}\geqq{10^{14}}}$\,GeV, \\
as shown in Fig.\,3.  
An extension to 5d GUTs \\ 
broken~\,by~orbifolds\,\cite{OB-GUT} \\
can~\,be~\,similarly~\,per-\\
formed\cite{5d3}.~The\,5d\,GUTs \\
with~nontrivial~UV\,fix- 
}}
\hspace{2.5mm} 
\lower125pt
\hbox{\epsfysize=9.3cm\epsfxsize=8.5cm\epsfbox{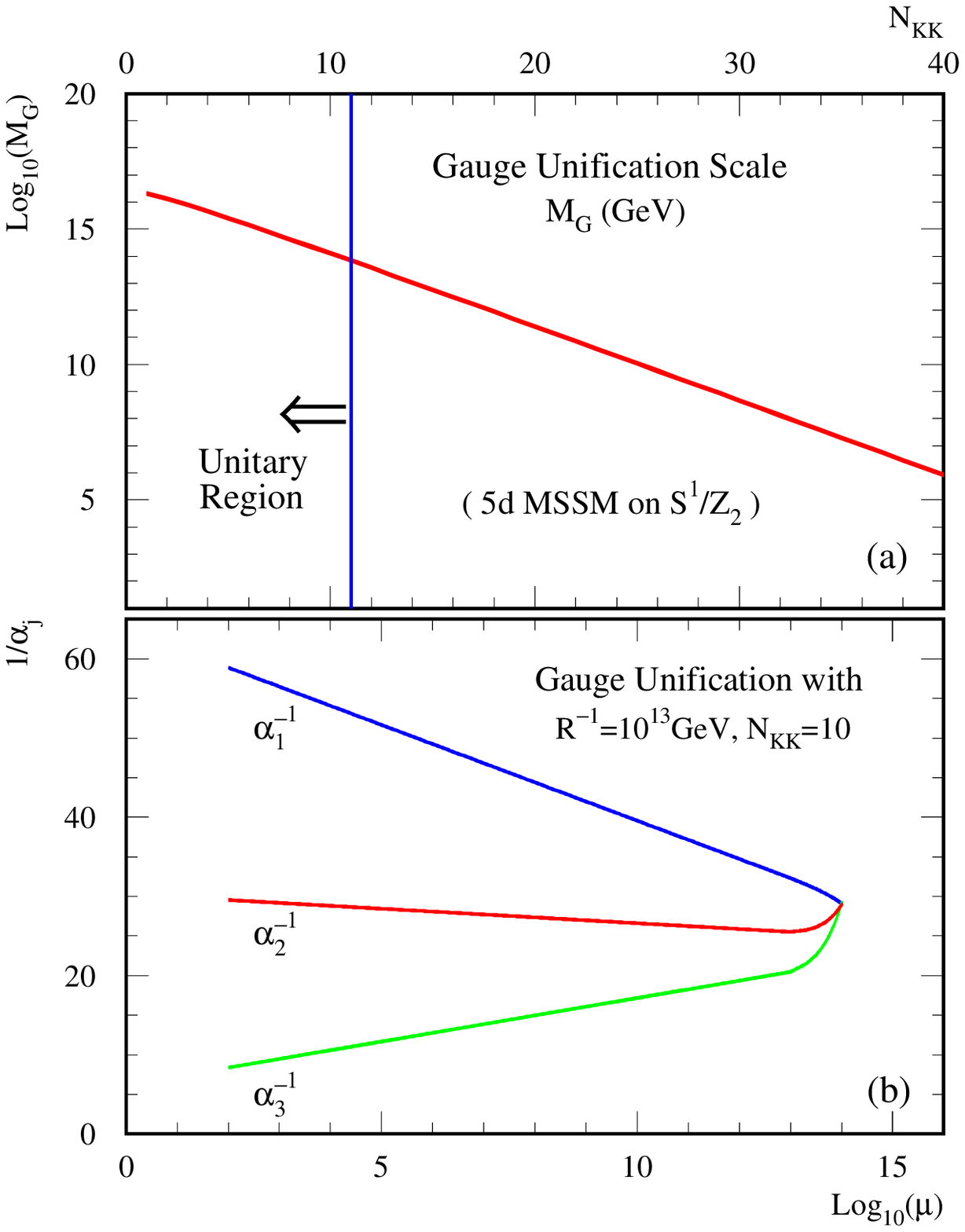}}
\end{math}
%

\vspace*{-3mm}
\hspace*{39mm}
{\small Figure\,3. The 5d SUSY gauge unification.}

\noindent
ed points\cite{UVF-GUT}
appear attractive to reconcile the unitarity constraint\cite{5d3}.

\vspace*{-3mm}

\section*{Acknowledgments}
\vspace*{-1.5mm}
H.J.H. thanks K. R. Dienes and T. Gherghetta for discussions at SUSY03,
and especially K. R. Dienes for his warm hospitality.
We also thank N. Arkani-Hamed, W. Bardeen, B. Dobrescu and C. T. Hill
for discussions.

\vspace*{-4.5mm}

\end{document}